\documentclass[aps,prd,twocolumn,showpacs,eqsecnum,nofootinbib]{revtex4}

\usepackage{epsfig}
\usepackage{graphicx}
\usepackage{dcolumn}
\usepackage{amsmath}
\usepackage{enumerate}
\usepackage{epstopdf} % remove if not Mac OS X

\newcommand{\trh}{T_{\rm RH}}

\begin{document}

\title{%
%\vskip-6pt \hfill {\rm\normalsize CERN-PH-TH/2008-052} \\
%\vskip-12pt~\\
Ultra-cold WIMPs: relics of non-standard pre-BBN cosmologies
}

\author{
\mbox{Graciela B.  Gelmini$^{1}$} and
\mbox{Paolo Gondolo$^{2}$}}

\affiliation{
\mbox{$^1$  Department of Physics and Astronomy, UCLA,
 475 Portola Plaza, Los Angeles, CA 90095, USA}
% \mbox{$^2$ CERN PH-TH, CH-1211, Geneva 23, Switzerland}
 \mbox{$^2$ Department of Physics, University of Utah,
   115 S 1400 E \# 201, Salt Lake City, UT 84112, USA }
\\
{\tt gelmini@physics.ucla.edu},
{\tt paolo@physics.utah.edu}}

%\date{\today}

\vspace{6mm}
\renewcommand{\thefootnote}{\arabic{footnote}}
\setcounter{footnote}{0}
\setcounter{section}{1}
\setcounter{equation}{0}
\renewcommand{\theequation}{\arabic{equation}}

\begin{abstract} \noindent
Weakly interacting massive particles (WIMPs) are one of very few probes 
of cosmology before Big Bang nucleosynthesis (BBN). We point out that 
in scenarios in which the Universe evolves in a non-standard manner 
during and after WIMP kinetic decoupling, the horizon mass 
scale at decoupling can be smaller and the dark matter WIMPs can be 
colder than in standard cosmology. This  would lead to much 
smaller  first objects in hierarchical structure  formation. In low 
reheating temperature scenarios the effect may be 
large enough as to noticeably enhance indirect detection signals in 
GLAST and other detectors, by up to two orders of magnitude.
\end{abstract}


\maketitle 

 In supersymmetric models, the lightest supersymmetric particle,
usually a neutralino $\chi$, is a good cold dark matter candidate.
 This is an example of a more general kind of cold dark matter candidates, 
 weakly interacting massive particles (WIMPs). The relic density and velocity distribution of WIMPs before structure formation depend on the characteristics of the Universe (expansion rate, composition, etc.) before Big Bang Nucleosynthesis (BBN), i.e.\ at temperatures  above $T\sim$ 1 MeV. This is an epoch from which we have no data. Indeed, if dark matter (DM)  WIMPs are ever found, they would be the first relics from that epoch that could be studied. Signatures of a non-standard pre-BBN cosmology that WIMPs may provide are few. Here we present one of, to our knowledge, only three.
  
In the standard  scenario of WIMP decoupling, one assumes that the entropy of matter and radiation is conserved,
that WIMPs are produced thermally, and, of most relevance for this work, that the first temperature of the radiation dominated epoch  before BBN is  high enough for WIMPs to have reached kinetic and
chemical equilibrium before they decouple. WIMPs decouple first chemically and then kinetically. The chemical decoupling (or freeze-out) temperature $T_{\rm fo}$ is the temperature after which their number practically does not change, and in the standard case $T_{\rm fo-std} \simeq m_{\chi}/20$, where $m_\chi$ is the WIMP mass.  The kinetic decoupling temperature $T_{\rm kd}$ is
the temperature after which WIMPs do not exchange momentum efficiently with the
cosmic radiation fluid. Within the standard cosmology (SC), $T_{\rm kd-std}$ lies between 10 MeV and a few GeV \cite{Profumo:2006bv}.

There are, however, well motivated cosmological models in which the standard assumptions above do not hold. These non-standard models include models with  gravitino~\cite{gravitino} or moduli~\cite{Moroi-Randall} decay, Q-ball decay \cite{Fujii}, 
thermal inflation \cite{Lazarides}, the  Brans-Dicke-Jordan~\cite{Brans-Dicke-Jordan}
cosmological model, models with anisotropic expansion~\cite{anisotropic} or quintessence domination~\cite{quintessence}. It has been pointed out that in all of these models the neutralino relic density $\Omega_\chi$ may differ from its standard value $\Omega_{\rm std}$ (see e.g.\ Ref.~\cite{Gelmini:2006pw}). 

One clear signature of a non-standard cosmology before BBN would be  WIMPs that compose part or all of the DM but would 
 be overabundant in the SC. Also the relic velocity distribution before structure formation in the Universe may differ from that in the SC. It has already been mentioned in the literature~\cite{Hisano:2000dz, Gelmini:2006vn} that WIMP's could be ``hotter" than in the SC, even constituting  warm instead of cold DM.  This would leave an imprint on the large scale structure spectrum.

Here we point out a third possible signature of non-standard pre-BBN cosmologies:  WIMPs may be  ``colder"  (i.e.\ they may have smaller typical velocities and thus smaller free-streaming length) and the  mass contained within the horizon at kinetic decoupling may be smaller than in the SC.
This would lead to a smaller mass for the smallest WIMP structures, those formed first. Some of the smallest WIMPs clumps would survive to the present.   Smaller and more abundant DM clumps 
 would be present within our galaxy, an observable consequence of which would be a stronger annihilation signal from our galactic halo detected in indirect DM  searches  by GLAST, PAMELA and other experiments~\cite{many, Bere-1, Green:2005fa, Loeb, Bertschinger:2006nq, Bere-2, Bere-3}. The signal in direct DM searches might also be affected in significant ways~\cite{Kamionkowski:2008vw}.     We show that  in low reheating temperature cosmological models. both the free-streaming mass scale  and the mass within the horizon at kinetic decoupling may be smaller than in the SC by many orders of magnitude 
  
We present estimates of the kinetic decoupling temperature, characteristic relic WIMP velocity, 
 free-streaming  and kinetic decoupling horizon mass scales using a generic  WIMP elastic scattering cross section written as in Ref.~\cite{Green:2005fa} and order of magnitude calculations.  After chemical decoupling, $T \lesssim T_{\rm fo}$, the total number of WIMPs remains constant and WIMPs are kept in local thermal equilibrium by elastic scattering with relativistic particles in the plasma. The WIMPs are non-relativistic at these temperatures, thus the average momentum exchanged per collision is small, of order $T$, and the rate of momentum exchange $\Gamma$ is suppressed by a factor $\sim T/m_\chi$ with respect to the rate of elastic scattering
\begin{equation}
\Gamma \equiv 
 \langle v \sigma_{\rm el}\rangle n_{\rm rad} \frac{T}{m_\chi}
 \simeq \sigma^{\rm el}_0 T^3 \left(\frac{T}{m_\chi}\right)^{2+l}~.
\label{gamma-1}
\end{equation}
Here $\sigma_{\rm el}$ is the total cross section for elastic
scattering of WIMPs and relativistic Standard Model fermions, $n_{\rm
rad} \simeq T^3$ is the number density of relativistic particles, which
are assumed to be in local thermal equilibrium, and $v\simeq 1$ is the WIMP-fermion relative velocity. The thermal
average of $\sigma_{\rm el}$ can be written as $\langle \sigma_{\rm
el}\rangle = \sigma^{\rm el}_0 (T/m_\chi)^{1+l}$, where $\sigma^{\rm el}_0
\simeq (G_{\rm F} m_{\rm W}^{2})^2 m^2/m_{\rm Z}^{4}\simeq 10^{-10} m_\chi^2$GeV$^{-4}$ sets the magnitude of the cross section, and $l$ parametrizes its temperature dependence.  Finally, $m_{\rm W}$ and $m_{\rm Z}$ are the masses of the
standard gauge bosons and $G_{\rm F}$ is Fermi's coupling constant.  In
the Standard Model, elastic scattering between a light fermion and a
heavy fermion is mediated by ${\rm Z}$ exchange and $l=0$. 
 In supersymmetric extensions of the
Standard Model, where the lightest neutralino is the WIMP candidate,
sfermion exchange occurs if the neutralino is a gaugino,
${\rm Z}$ exchange is suppressed, and $l=1$. Inserting $\sigma^{\rm el}_0$ into Eq.~\ref{gamma-1}, the  rate of momentum exchange for non-relativistic WIMPs is
\begin{equation}
\Gamma  \simeq \frac{10^{-10} T^5}{{\rm GeV}^4} \left(\frac{T}{m_\chi}\right)^l.
\label{gamma-2}
\end{equation}
In the following, we focus on the case of neutralinos with $l=1$ (analogous results can be easily derived for $l=0$). 

Kinetic decoupling occurs when the rate of momentum exchange becomes smaller than the expansion rate of the Universe $H$. In the SC, decoupling happens while the universe is radiation dominated so the Hubble parameter is 
$H \simeq T^2/M_P$, where $M_P \simeq 10^{19} {\rm GeV}$ is the Planck scale. From $\Gamma \simeq H$, we get
\begin{equation}
T_{\rm kd-std} 
 \simeq 20 {\rm MeV}
  \left(\frac{m_\chi}{100 {\rm GeV}}\right)^{1/4}.
\label{Tkd-std}
\end{equation}
More accurate calculations give a range of 10 MeV to a few GeV for $T_{\rm kd-std}$~\cite{Profumo:2006bv}.

We concentrate now on a class of non-standard cosmological models with a late episode of inflation or entropy production~\cite{gravitino, Moroi-Randall, Lazarides, Gelmini:2006pw} in which  a scalar field $\phi$ dominates the energy density of the Universe and subsequently decays (while oscillating around a minimum of its potential) eventually reheating the Universe to a low reheating temperature $\trh$. This does not spoil primordial nucleosynthesis provided $\trh \gtrsim$4 MeV~\cite{hannestad}. The interesting case for us is when $\trh$ is smaller than the standard chemical decoupling temperature $T_{\rm fo-std}$, so kinetic decoupling happens during the $\phi$-oscillations dominated phase.

 Late-decaying scalar field models are well motivated in particle theories. For example,
moduli fields which acquire a mass $m_\phi$ at the supersymmetry breaking scale 10 to 100 TeV and have gravitational strength interactions, thus their decay rate is 
$\Gamma_{\rm decay} \simeq m_\phi^3/ M_p^{2}$,  are pervasive in supersymmetric models. These fields naturally tend to dominate the energy density of the Universe at late times and produce reheating temperatures in the MeV range (the ``moduli problem"~\cite{moduli-problem}  is the tendency of these decays to happen even after BBN, which must be avoided). In fact,  approximating the decay as instantaneous (usually a very good approximation) at the moment of decay the energy stored in the field goes into radiation at a temperature $T_{\rm RH}$. Thus $\Gamma_{\rm decay} \simeq 
H(T_{\rm RH}) \simeq T_{\rm RH}^2/ M_P$ emplies that 
\begin{equation} 
T_{\rm RH}\simeq 10~{\rm MeV}\left(\frac{m_\phi}{\rm
100~TeV}\right)^{3/2} .
\label{TRH}
\end{equation}

During the epoch in which the Universe
is dominated by the oscillating $\phi$ field,  the Hubble parameter $H_\phi$ is proportional to
$T^4$~ \cite{McDonald}. Since at the moment of $\phi$ decay, when $T=\trh$,  
$H_\phi (\trh) \simeq \trh^2/M_P$, we can fix the proportionality constant so that  $H_\phi \simeq T^4/(\trh^2 M_P)$.  Requiring that  $\Gamma \simeq H_{\phi}$ at the new kinetic decoupling temperature $T_{\rm kd'}$, we obtain
\begin{equation}
T_{\rm kd'}
 \simeq 30 {\rm MeV}  \left(\frac {10 {\rm MeV}}{T_{\rm RH}}\right)
  \left({\frac{m_\chi} {100 {\rm GeV}}}\right)^{1/2}~.
\label{Tkd'}
\end{equation}
Combining the two equations $\Gamma(T_{\rm kd-std}) \simeq T_{\rm kd-std}^2/M_P$ and $\Gamma(T_{\rm kd'}) \simeq T_{\rm kd'}^4/(\trh^2
M_P)$, we have
\begin{equation}
T_{\rm kd'} \simeq \frac{T_{\rm kd-std}^2}{T_{\rm RH}}~.
\label{Tkd'-2}
\end{equation}
Thus, if the reheating temperature is smaller than the standard kinetic decoupling temperature ($T_{\rm RH}<T_{\rm kd-std}$), WIMPs decouple earlier than in the SC, i.e.\ $T_{\rm kd'} > T_{\rm kd-std}$,   and do so during the $\phi$-oscillations dominated epoch.

So far we have assumed that the WIMPs are non-relativistic at decoupling. Thus, the relations above hold for  $T_{\rm kd'} < m_\chi/3$. If Eq.~\ref{Tkd'-2} leads to $T_{\rm kd'} > m_\chi/3$, the WIMPs would be relativistic at decoupling, and the equations need to be modified. Since the momentum transfer to the radiation background is very efficient in collisions of relativistic WIMPs, in general the kinetic decoupling would happen at the moment WIMPs become non-relativistic and not earlier, i.e.\ $T_{\rm kd'} \simeq m_\chi/3$ (unless the scattering cross section is so small that WIMPs are never in kinetic equilibrium~\cite{Gelmini:2006vn}).

Late-decaying scalar field models exemplify many combinations of reheating and decoupling temperatures. In these models, the dominant WIMP production mechanism can be thermal (due to interactions with the radiation background) or non-thermal (due to the decay of the $\phi$ field into WIMPs), with or without chemical equilibrium (see for example Ref.~\cite{Gelmini:2006pw}). 
In these models, neutralinos in almost all supersymmetric models could have the relic density necessary to be the DM~\cite{Gelmini:2006pw} through a combination of thermal and non-thermal production mechanisms~\cite{Moroi-Randall,kamionkowski-turner,Moroi,Chung,Giudice,Drees,Khalil,Fornengo,Pallis}.
For thermal production without chemical equilibrium, most WIMP production happens at $T_\star \simeq m_\chi/4$~\cite{Chung}. Thus the WIMP number per comoving volume is fixed then. For $T_{\rm kd'} < T_\star$, the equations we derived above hold. 
For thermal production with chemical equilibrium the neutralino freezes out while the Universe is dominated by the $\phi$ field at a new freeze-out temperature $T_{\rm fo'}$ higher than the usual $m_\chi/20$~\cite{McDonald,Giudice}. The freeze-out density is larger than usual, but it is diluted by entropy production from
$\phi$ decays, $\Omega_\chi \simeq \trh^3 T_{\rm fo-std} (T_{\rm fo'})^{-4} \Omega_{\rm std}$.
The numerical results in Ref.~\cite{Gelmini:2006pw}
indicate a dependence of $\Omega_\chi$  closer to $\trh^4$, and that  the freeze-out temperature depends of $T_{\rm RH}$ as $T_{\rm fo'} 20/m_\chi \simeq (T_{\rm RH} 20/m_\chi)^{1/4}$. In this case, the equations we derived hold for $T_{\rm kd'} < T_{\rm fo'}$.
Non-thermal production without chemical equilibrium happens when production of WIMPs in the decay of the $\phi$ field is not compensated by annihilation. 
WIMPs are produced with an energy which is a fraction $f$ of the $\phi$-field mass $m_\phi$, $E_\chi \simeq f m_\phi$. Thus if $m_\chi < 3 f m_\phi$, WIMPs are produced relativistic. In this case the elastic scattering cross section  is $\langle v \sigma_{\rm el}\rangle \simeq  \sigma_0 T E_\chi/ m_\chi^2\simeq \sigma_0 T f m_\phi/ m_\chi^2$.  Taking the characteristic value of $\sigma_0$ as above, neutralinos are in kinetic equilibrium  while they are relativistic for $m_\phi> 10~{\rm keV}/ f (T_{\rm RH}/10~ {\rm MeV})^2$, that is for all  the physically acceptable values of $m_\phi$. 
 \begin{figure}[t]
\includegraphics[width=3.3in]{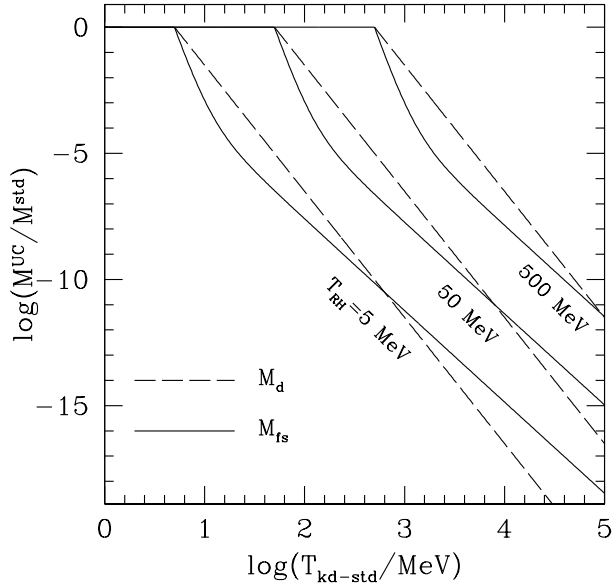}
\caption{Ratios of the late decaying scalar field (with ultra cold (UC) WIMPs) and standard (std) scenario free-streaming scales ($M^{\rm UC}_{\rm fs}/M^{\rm std}_{\rm fs}$, solid lines) and damping scales ($M^{\rm UC}_{\rm d}/M^{\rm std}_{\rm d}$, dashed lines), as functions of the standard kinetic decoupling temperature $T_{\rm kd-std}$. Each set of two lines is labeled by the corresponding value of the reheating temperature $T_{\rm RH}$.}
\end{figure}
 Thus kinetic decoupling occurs while neutralinos are non-relativistic as assumed above. In the extreme case in which neutralinos are never in kinetic equilibrium~\cite{Gelmini:2006vn}, neutralinos can actually be warm DM \cite{Hisano:2000dz}, since they are produced late in the history of the Universe and with a large initial energy that redshifts  until the moment of structure formation.

We now estimate the WIMP characteristic speed in low $T_{\rm RH}$ models.  At the moment of kinetic decoupling, WIMPs are in thermal equilibrium with the radiation, and their characteristic speed is
$v'(T_{\rm kd'}) \simeq \sqrt{T_{\rm kd'}/m_\chi}$. After decoupling, the speed decreases as $a^{-1}$ (if the WIMPs are non-relativistic).  During the $\phi$-oscillations dominated era, the scale factor of the Universe is related to the radiation temperature as $a \sim T^{-8/3}$. When the Universe becomes radiation dominated, i.e.\ at $T_{\rm RH}$, the characteristic speed of WIMPs is therefore
\begin{equation}
v'(T_{\rm RH}) \simeq \sqrt{{T_{\rm kd'}}/{m_\chi}} \left( {T_{\rm RH}}/{T_{\rm kd'}} \right)^{8/3}.
\label{v'}
\end{equation}
In comparison, in the standard radiation dominated case, $a \sim T^{-1}$ and at the same temperature $T_{\rm RH}$
WIMPs have a characteristic speed
\begin{equation}
v_{std}(T_{\rm RH}) \simeq \sqrt{{T_{\rm kd-std}}/{m_\chi}} \left({T_{\rm RH}}/{T_{\rm kd-std}}\right).
\label{v}
\end{equation}
Because speeds redshift in the same way in both models at temperatures smaller than $T_{\rm RH}$, after reheating the speeds $v'$ and $v_{std}$ remain in the same ratio 
\begin{equation}
{{v'}/{v_{std}}} \simeq  \left({T_{\rm RH}}/{T_{\rm kd-std}}\right)^{10/3}.
\label{ratiov'/v}
\end{equation}
This relation applies to the case  $T_{\rm kd'} < m_\chi/3$ for which Eq.~\ref{Tkd'-2} holds.
Thus the characteristic relic WIMP speed in low $T_{\rm RH}$ cosmological models can be much smaller than in the SC. In other words, WIMPs can be much colder, i.e.  ``ultra-cold", as we call them. 

The free-streaming length $\lambda_{\rm fs}$ of ultra-cold WIMPs is consequently smaller than that of standard WIMPs. $\lambda_{\rm fs}$ is the characteristic distance covered by WIMPs from the time of kinetic decoupling $t_{\rm kd}$ to the present (while they propagate as free particles) 
\begin{equation}
\lambda_{\rm fs}= c a_0 \int_{t_{\rm kd}}^{t_{0}} v \frac{dt}{a} 
\simeq c \sqrt{\frac{T_{\rm kd}}{m_\chi}}a_0 a_{\rm kd} \int_{a_{\rm kd}}^{a_{0}} \frac{da}{a^3 H(a)}.
\label{lambda-general}
\end{equation}
Here, $T_{\rm kd}$ and $a_{\rm kd} $ are the temperature and scale factor at the moment of kinetic decoupling, and $a_0$ is the value of the scale factor at present.  During the $\phi$-oscillations dominated epoch, $H(a) \propto a^{-3/2}$ with $a \sim t^{2/3} \sim T^{-8/3}$ and $H \simeq T^4/(T_{\rm RH}^2 M_P)$.   During the radiation dominated epoch, $H(a) \propto a^{-2}$, with $a \sim t^{1/2} \sim T^{-1}$ and $H \simeq T^2/ M_P$, and the integral in the definition of $\lambda_{\rm fs}$ is $\propto \ln a$. During the matter-dominated epoch, $H(a) \propto a^{-3/2}$, with $a \sim t^{2/3} \sim T^{-1}$ and $H \simeq T^{3/2}/ M_P$, and the free-streaming length saturates.

When the kinetic decoupling occurs during the $\phi$-oscillations dominated epoch and WIMPs are ultra-cold (UC), we obtain
\begin{eqnarray}
\lambda^{\rm UC}_{fs} & \simeq & c \sqrt{\frac{T_{\rm kd'}}{m_\chi}} \frac{a_{\rm kd'} M_P}{a_0 T_0^2}
\left\{ 2 \left[ \left(\frac{T_{\rm kd'}}{T_{\rm RH}} \right)^{4/3} -1 \right] + \right.\nonumber\\ && \hspace{1.1in} \left.
\ln{\left(\frac{T_{\rm  RH}}{T_{\rm eq}}\right)} + 1.96 \right\} .
\label{lambda-ours-1}
\end{eqnarray}
The first term within the curly brackets arises from the $\phi$-dominated epoch, the logarithmic term from the radiation dominated epoch, and the term $ 1.96=2[1-(1+z_{\rm eq})^{-1/2}]$ from the matter dominated epoch. The subindex ${\rm eq}$ refers to matter-radiation equality.
When Eq.~\ref{Tkd'-2} holds (for $T_{\rm kd'} < m_\chi/3$), Eq.~\ref{lambda-ours-1} becomes
\begin{eqnarray} 
&& \lambda^{\rm UC}_{\rm fs}\simeq \frac{c M_P}{T_0 \sqrt{m_\chi T_{\rm kd-std}}} 
\left(\frac{T_{\rm RH}}{T_{\rm kd-std}}\right)^{23/6} \times \nonumber \\~\nonumber\\
& & \left\{ 2 \left[ \left(\frac{T_{\rm kd-std}}{T_{\rm RH}} \right)^{8/3} -1 \right] + 
\ln{\left(\frac{T_{\rm  RH}}{T_{\rm eq}}\right)} +1.96 \right\}.
\quad
\label{lambda-ours-2}
\end{eqnarray}
This is to be compared with the  free-streaming length in the SC,
\begin{equation}
\lambda^{\rm std}_{\rm fs} \simeq  \frac{c M_P}{T_0 \sqrt{m_\chi T_{\rm kd-std}}}
 \left[ \ln{\left(\frac{T_{\rm kd-std}}{T_{\rm eq}}\right)} + 1.96 \right].
\label{lambda-usual}
\end{equation}
As traditional, we introduce the mass $M_{\rm fs}$ contained within a sphere of radius $\lambda_{\rm fs}/2$, and compare the standard and non-standard scenarios through the ratio
\begin{equation}
 \frac{M^{\rm UC}_{\rm fs}}{M^{\rm std}_{\rm fs}} = \left( \frac{\lambda^{\rm UC}_{\rm fs}} {\lambda^{\rm std}_{\rm fs}} \right)^3. 
 \end{equation}
As shown in Fig.~1 (solid lines), this ratio is always smaller than 1 if the reheating temperature is smaller than the standard kinetic decoupling temperature, and it can be many orders of magnitude smaller.
For $T_{\rm RH} =5$ MeV and $T_{\rm kd-std}$ between 10 MeV and a few 
GeV~\cite{Profumo:2006bv}, the 
free-streaming scale $M_{\rm fs}$ can decrease by a factor between 0.1 and  10$^{-13}$ (the suppression is less important for larger values of $T_{\rm RH}$).

Friction between WIMPs and relativistic leptons during kinetic decoupling (Silk damping) leads to a small-scale cutoff in structure formation at the scale of the horizon at kinetic decoupling~\cite{Loeb,Bertschinger:2006nq}.  The mass contained within the horizon at decoupling in the SC  is $M_d^{\rm std} \simeq 10^{-4} M_\odot ($10 MeV$/T_{\rm kd-std})^3$~\cite{Loeb, Bertschinger:2006nq}. It varies from $10^{-4} M_\odot$ for $T_{\rm kd-std} \simeq 10$ MeV to
 $10^{-12} M_\odot$ for $T_{\rm kd-std} \simeq 5$ GeV . 
 
 During the $\phi$-oscillations dominated phase the Universe expands, and thus the density contrast of DM inhomogeneities grow,  in the same way as in a matter dominated phase. A detailed study (which is beyond the scope of this paper) of  the kinetic decoupling during this phase should be done to find the cut-off mass scale of the smallest dark matter structures, which will also depend on the particular particle physics model considered.  However, it is reasonable to
assume that also in this case the cut-off will be given by the comoving free-streaming mass scale and/or  the kinetic decoupling horizon mass scale.
 
During the $\phi$-oscillations dominated phase the matter density scales as $T^{-8}$ and the time  as $T^{-4}$. Thus,  the mass contained in the horizon at decoupling $M_d^{\rm UC}$ is smaller than in the SC by the factor
\begin{equation}
\frac{M_d^{\rm UC}}{M_d^{\rm std}} \simeq \left(\frac{T_{\rm RH}}{T_{\rm kd'}} \right)^4\left(\frac{T_{\rm kd-std}}{T_{\rm RH}} \right)^3.
 \label{ratioMd}
\end{equation}
Using Eq.~\ref{Tkd'-2}, this ratio becomes $(T_{\rm RH}/ T_{\rm kd-std})^5$. As seen in Fig.~1 (dashed lines), the suppression factor $(M_d^{\rm UC}/M_d^{\rm std})$ can be substantial, ranging from 0.1 for $T_{\rm kd-std} = 10$ MeV to  10$^{-15}$ for $T_{\rm kd-std} = 5$ GeV, when assuming  $T_{\rm RH} =5$ MeV (the suppression is less important for larger values of  $T_{\rm RH}$). This means that the range of $M_d^{\rm UC}$ is now from  $10^{-27} M_\odot$ for $T_{\rm kd-std} \simeq 5$ GeV to  $10^{-5} M_\odot$ for $T_{\rm kd-std} \simeq 10$ MeV. 

 From the ratios $(M_d^{\rm UC}/M_d^{\rm std})$ and $(M_{\rm fs}^{\rm UC}/M_{\rm fs}^{\rm std})$ shown in Fig.~1, we see that the damping mass scale $M_{d}$ is less  suppressed than the free-streaming mass scale $M_{\rm fs}$, except possibly for standard kinetic decoupling temperatures above 1 GeV.  This is important because  the scale of the smallest WIMP haloes will be the larger of  $M_{\rm fs}$ and  $M_{d}$, since a halo mass must be larger than both.  In the standard cosmological scenario, the damping scale $M_{d}^{\rm std}$ is usually larger than the free-streaming scale (by a factor $(m_\chi/T_{\rm kd-std})^{3/2}$, at least for supersymmetric gaugino models with $T_{\rm kd-std}$ in the 10--100 MeV range~\cite{Loeb,Bertschinger:2006nq}).  Fig.~1 shows that in low $T_{\rm RH}$ models the damping scale remains larger than the free-streaming scale for most typical values  of the standard kinetic decoupling temperature  $T_{\rm kd-std} \lesssim$1 GeV. Thus, we take the damping scale $M_{d}$ to be the characteristic mass of the smallest WIMP halos. 
 
If the smallest halos survive until today, they may be present in the dark halo of our galaxy and may enhance the expected WIMP annihilation signals over the smooth halo expectation by a boost factor $B$. The $B$ factor increases slowly with decreasing $M_d$. For a halo of mass $M$ and  smallest subhalo mass $M_d$, Ref.~\cite{B} finds $B \simeq 0.1 [(M/M_d)^{0.13}-1]$.  In the SC, one expects $B$ of the order of 10 for the Milky Way  for which $M\simeq 10^{12} M_\odot$. For example, from the equation just mentioned we get $B \simeq 20$  to 130 for  the standard range of $M_d$, from $10^{-6} M_\odot$ to $M_d \simeq 10^{-12} M_\odot$.  With ultra-cold WIMPs,  $M_d$ could be much smaller and thus the boost factor could be much larger: it could reach $B \simeq 10^{4}$  for  $M_d\simeq10^{-27} M_\odot$.  Such large boost factors would not only make a halo WIMP annihilation signal easier to detect, but would also be a signature of a non-standard pre-BBN cosmology.

We finally remark that ultra-cold WIMPs may arise in all models in which the expansion rate of the Universe in the pre-BBN era is larger than assumed in the SC, although the magnitude of the effect would in general be  smaller than in the low reheating temperature models presented above. For example, let us consider ``kination" models~\cite{quintessence}. These are models in which the kinetic energy of a scalar field, 
$\rho_\phi = \dot {\phi}^2/ 2 \sim a^{-6}$, 
dominates the energy density of the Universe at $T>T_{\rm kin}$ before BBN, while the entropy is dominated by the radiation, thus $a \sim T^{-1}$. Roughly, $\rho_\phi /  \rho_{\rm rad} \simeq \eta_\phi (T/{\rm MeV})^2$, where $\eta_\phi$ is the fraction of the energy density of the Universe at $T \simeq$ 1 MeV due to the kinetic energy of the $\phi$ field. At higher temperatures the fraction of $\phi$ kinetic energy grows very fast with the temperature and it is dominant for $T> T_{\rm kin}\simeq \eta^{-1/2}_\phi$ MeV. If the kinetic decoupling of WIMPs happens during the 
kination period, i.e. if $T_{\rm kd-std} >
 T_{\rm kin}$, assuming $H \simeq \sqrt{\rho_\phi}/ M_P$, the kinetic decoupling happens approximately  at
\begin{equation}
T_{\rm kd'}^{\rm kin} \simeq  50 {\rm MeV} \eta_\phi^{1/6}
  \left({\frac{m_\chi} {100 {\rm GeV}}}\right)^{1/3}.
\label{Tkd'-kin}
\end{equation}
In this case the free-streaming length is
\begin{eqnarray}
\lambda^{\rm kin}_{\rm fs} & \simeq & c \sqrt{\frac{T_{\rm kd'}}{m_\chi}} \frac{a_{\rm kd'} M_P}{a_0 T_0^2}
\left\{ 1 - \frac{T_{\rm kin}}{T_{\rm kd'}^{\rm kin}}+ \right.\nonumber\\ && \hspace{1.1in} \left.
\ln{\left(\frac{T_{\rm kin}}{T_{\rm eq}}\right)} + 1.96 \right\},
\label{lambda-ours-kinetion}
\end{eqnarray}
which for $T_{\rm kd'}^{\rm kin}> T_{\rm kd-std} > T_{\rm kin}$ is smaller than the standard free-streaming length. During the kination period,  the scale factor is  $a \sim T^{-1}$ and the time evolves as $t \sim T^{-3}$ thus the mass contained within the horizon at kinetic decoupling $M_d^{\rm kin}$,  again for $T_{\rm kd'}^{\rm kin}> T_{\rm kd-std} >
 T_{\rm kin}$, is smaller than the standard mass scale $M_d^{\rm std}$
by the ratio
\begin{equation}
\frac{M_d^{\rm kim}}{M_d^{\rm std}} \simeq \frac{T_{\rm kin}^3  T_{\rm kd-std}^3}{\left(T_{\rm kd'}^{\rm kin}\right)^6}.
 \label{ratioMdkination}
\end{equation}
Thus in kination models the free-streaming and damping mass scales can be smaller than in the SC,  although not by as much as in the late decaying scalar  field (or low reheating temperature) models presented above.

We have here pointed out that a too large boost factor in the annihilation signal of a particular WIMP would be a signature of a non standard cosmological evolution of the Universe just before BBN, during the kinetic decoupling of WIMPs.  If dark matter WIMPs are ever found, they would be the first relics from the pre-BBN epoch that could be studied. Signatures of a non-standard pre-BBN cosmology that WIMPs may provide are few and here we presented one of them:  WIMPs may be ultra cold  so the mass of the smallest WIMP structures, those formed first, may be smaller than in the standard cosmology. Some of the smallest WIMPs clumps would survive to the present.   Smaller and more abundant DM clumps 
 would be present within our galaxy, an observable consequence of which would be a stronger annihilation signal from our galactic halo detected in indirect DM  searches  by GLAST, PAMELA and other experiments.  Boost factors as large as $10^4$ for usual WIMP candidates are possible in the low reheating temperature scenarios considered here.  In last instance, verifying  this signature would require to study  in accelerators, the LHC or ILC,  the properties of the particular WIMP  that would allow us to estimate its scattering cross section, to find the same WIMP in indirect detection searches and to understand the formation and survival of  the earliest dark matter clumps within the halo of our galaxy. 
  
This work was supported in part by the US Department of Energy Grant
DE-FG03-91ER40662, Task C  at UCLA, and NSF
grant PHY-0456825 at the University of Utah.

\end{document}